# A Comprehensive Guide to Item Recovery Using the Multidimensional Graded Response Model in R

Yesim Beril Soguksu[1], Ayse Bilicioglu Gunes[2], Hatice Gurdil[3]

[1]Ministry of National Education, Kahramanmaraş, Türkiye; [2]Bartın University, Department of Measurement and Evaluation in Education, Bartın, Türkiye; [3]Ministry of National Education, Ankara, Türkiye

**Abstract**

The purpose of this study is to provide a step-by-step demonstration of item recovery for the Multidimensional Graded Response Model (MGRM) in R. Within this scope, a sample simulation design was constructed where the test lengths were set to 20 and 40, the interdimensional correlations were varied as 0.3 and 0.7, and the sample size was fixed at 2000. Parameter estimates were derived from the generated datasets for the 3-dimensional GRM, and bias and Root Mean Square Error (RMSE) values were calculated and visualized. In line with the aim of the study, R codes for all these steps were presented along with detailed explanations, enabling researchers to replicate and adapt the procedures for their own analyses. This study is expected to contribute to the literature by serving as a practical guide for implementing item recovery in the MGRM. In addition, the methods presented, including data generation, parameter estimation, and result visualization, are anticipated to benefit researchers even if they are not directly engaged in item recovery.

**Introduction**

Classical Test Theory (CTT; Spearman, 1904) has been extensively employed in the development of scales and tests over several decades. However, efforts to address the limitations of this theory led to the establishment of the foundations of Item Response Theory (IRT) by Lord (1952) and Birnbaum (1968). In IRT, responses to items are modeled as a result of the interaction between item parameters and individuals' ability levels (Meijer & Tendeiro, 2018). A key advantage of IRT is its capacity to estimate individuals' abilities independently of the items and to determine the psychometric properties of the items independently of the sample (Embretson & Reise, 2000; Lord, 2012). In Unidimensional IRT models, the accurate estimation of item parameters and the validity of these models in following applications depend on meeting the unidimensionality assumption of the data. Unidimensionality is defined as the presence of a single latent trait underlying the data (Hattie, 1985). This assumption posits that variability in item responses is attributable to a single latent variable. However, in practice, achieving strict unidimensionality is uncommon, as psychological constructs often encompass complex, multidimensional traits. In responding to test items, individuals may rely on multiple abilities or skills simultaneously (Reckase, 2009). The challenge of ensuring unidimensionality has catalyzed the exploration of multidimensional constructs, underscoring the necessity for measurement models capable of capturing the multidimensional nature of psychological attributes with greater precision. This demand has led to the development of Multidimensional Item Response Theory (MIRT).

MIRT is a statistical model that describes the relationship between two or more latent variables, referred to as dimensions, and the probability of correctly responding to a given test item (Ackerman et al., 2003). By allowing for a more accurate modeling of performance involving complex structures, MIRT represents a significant advancement in the field of measurement. Particularly in educational settings, it is a highly relevant model, as most tasks often require the engagement of multiple abilities (Hartig & Höhler, 2009). In unidimensional models, the probability of correctly responding to an item is determined by a single dominant ability dimension. In contrast, MIRT models this probability as a function of multiple ability dimensions. MIRT models are further categorized into compensatory and non-compensatory types, depending on whether high proficiency in one dimension can offset low proficiency in another (Sijtsma & Junker, 2006). In compensatory MIRT models, latent abilities interact in such a way that a deficiency in one ability can be offset by a higher level of another ability. Conversely, in non-compensatory MIRT models, each measured ability must meet a sufficient threshold, and a deficiency in one ability cannot be fully compensated by strengths in other abilities. Beyond their capacity to model multidimensional constructs, MIRT models have increasingly been extended to a broader range of applications. This expansion is particularly notable in the transition from dichotomously scored items to polytomously scored items.

Meijer and Tendeiro (2018) state that IRT models were initially developed for dichotomously scored items due to their having a simpler theoretical structure, but these models were subsequently extended to accommodate polytomous items. Furthermore, although polytomously scored items present a more complex structure, their use has increased significantly in place of dichotomous items due to the growing emphasis on assessing higher-order skills. This shift is largely driven by the ability of polytomous items to provide more detailed information compared to dichotomous items (Thissen, 1976). Moreover, polytomously scored items have long been preferred not only for measuring cognitive traits but also for assessing affective traits. In response to this need, polytomous IRT models were developed (Samejima, 1969; Bock, 1972; Andrich, 1978).

Different from dichotomously scored IRT models, the unit of analysis in polytomous IRT models is not the item but the response categories. Each response category has its own response function (CRF) and the model is analyzed through these functions. Existing polytomous IRT models, which were initially developed for unidimensional constructs, have been extended to multidimensional, polytomous IRT models thereby enabling the examination of constructs that involve more than one dimension or latent trait (Muraki & Carlson, 1995; Yao & Schwarz, 2006). Ackerman (1996) highlighted that most psychological and educational tests measure multiple traits or combinations of traits to varying degrees. Multidimensional and polytomous IRT models enable the evaluation of multiple dimensions by allowing each item to be rated across more than one level, thereby facilitating a more nuanced assessment. These models provide more precise and comprehensive analyses by considering the performance of each dimension at different levels, effectively capturing students' performance across various skills and knowledge areas, particularly in educational assessments.

IRT, which encompasses a variety of models suitable for different data structures, is often utilized in Monte Carlo (MC) simulation studies to examine the accuracy of parameter estimation and evaluate model performance (Alarcon, Lee & Johnson, 2023; Bilicioglu Gunes, 2024; Furlow, Ross & Gagné, 2009; Gurdil et al., 2024; Walker & Gocer Sahin, 2016). MC is highly popular among researchers due to its ability to test models under various scenarios. It offers several advantages, including ease of use, flexibility, efficiency, the utilization of random assignment, rapid results, and the capability to evaluate models under diverse conditions with large datasets (Kroese et al., 2014). Emerging as a popular approach in the late 19th and early 20th centuries, MC studies facilitate the use of diverse data structures by addressing challenges such as limited access to real datasets. This method allows researchers to generate data for evaluating models under various conditions, enabling analyses with different parameters that adequately address their research questions (Hallgren, 2013). In the field of psychometrics, MC simulation is widely used to answer a wide range of research questions, from evaluating the success of various models under different conditions (e.g., small sample conditions) to the robustness of the adaptability of unidimensional models to multidimensional data, especially in the context of IRT (Harwell et al., 1996). It is also used professionally in some cases for routine tasks such as the evaluation of model data (Hambleton et al., 1991).

In MC simulations, determining the simulation conditions and defining the replication process are critical steps. When establishing simulation conditions, a thorough review of the literature on the research topic is essential. The conditions should be designed to reflect scenarios commonly addressed in the literature or those likely to occur in real-world applications. Moreover, as noted by Feinberg and Rubright (2016), careful consideration must be given to the number of replications to ensure accurate and reliable parameter estimation. However, the critical question remains: how many replications are sufficient? Data generated with a limited number of replications may lack robustness and compromise the reliability of the results. On the other hand, using an excessive number of replications may introduce challenges such as extended computation time or limitations of the hardware utilized. Harwell et al. (1996) recommended a minimum of 25 replications for IRT studies; however, the appropriate number of replications should ultimately be determined based on the specific context of the research problem. At this stage, it is considered more pragmatic to make this decision by taking into account practical constraints, such as model complexity, hardware and software limitations, the number of conditions, and time availability (Feinberg & Rubright, 2016).

In MC simulations, various software programs can be utilized for data generation; nonetheless, the R programming language stands out as a popular choice among researchers due to its open-source nature, cost-free accessibility, and extensive range of available packages. However, researchers, particularly those less experienced with the R programming language, may encounter challenges in generating multidimensional and polytomous data.

This limitation can hinder researchers lacking sufficient technical expertise from producing complex datasets across various dimensions.

A review of the literature highlights several simulation studies based on MIRT models (Gurdil & Demir, 2024; Han & Paek, 2014; Strachan, 2020; Xu et al., 2017; Zhang, 2012). These studies examined the effects of various configurations of MIRT models, particularly focusing on model fit statistics, calibration accuracy, and model robustness. They also investigated the effectiveness of MIRT in smoothing varying item response functions and compared the performance of different software packages. Similarly, simulation studies on the Multidimensional Graded Response Model (MGRM) (Jiang et al., 2016; Kehinde et al., 2022; Reise & Yu, 1990) have focused on factors influencing item recovery. Additionally, Bulut and Sünbül (2017) provided a comprehensive guide on detecting Differential Item Functioning (DIF) in MGRM using R.

In contrast to previous studies, this study aims to offer a detailed, step-by-step guide for performing item recovery in MGRM using R, including the R codes required for each step. Specifically, the study will provide clear explanations on data generation, parameter estimation, RMSE and bias calculations, and result visualization, ensuring accessibility for researchers with varying levels of technical expertise. The study is expected to make valuable contributions, particularly to researchers working with MIRT models and conducting Monte Carlo simulations using R. Moreover, even researchers not directly involved in item recovery can benefit from this work for specific tasks such as data generation, parameter estimation, or visualization.

The subsequent sections are organized as follows: The 'Method' section describes the simulation design, including the levels of conditions and parameters along with their theoretical justifications. It also explains the data generation process based on the MGRM. The next sections detail the parameter estimation procedures, the calculations of bias and RMSE, and the visualization of the findings. Finally, the 'Discussion' section critically evaluates the study's outcomes.

**Method**

This section provides detailed information on the process of data generation for the multidimensional GRM using the R programming language, the estimation of parameters, the calculation of bias and RMSE values for these parameters, and the visualization of results through graphical representations. To this end, a sample simulation design was first developed based on insights from the literature. Subsequently, the R packages and functions employed were introduced, and the relevant procedures were outlined step by step. The study was conducted using R version 4.3.3.

*Simulation Design*

Data generation for MC simulation was performed according to GRM (Samejima, 1969), which is one of the most widely used MIRT models (Carle et al., 2009; French & Vo, 2020; Kuo & Sheng, 2015; Sharkness & DeAngelo, 2011; Uttaro & Lehman, 1999). GRM is specifically designed for data with ordered categories. Its ability to estimate threshold parameters for each category allows it to model ordered data structures with greater precision. This distinctive feature of GRM enables more accurate parameter estimation, particularly in the analysis of ordered responses, compared to other polytomous models. Furthermore, its capacity to deliver stable results in the context of large sample sizes and extended test lengths makes it a preferred choice in simulation studies (De Ayala, 2009).

A three-dimensional MGRM model was employed in this study, as it enhances the measurement power of the test by more accurately capturing the skills assessed by the items and the interrelationships among these skills (Adams et al., 1997). Two-dimensional models, while useful, may fall short in adequately revealing the relationships between dimensions and may limit the precision of parameter estimates. In contrast, a three-dimensional structure addresses these limitations, offering a more realistic representation of the constructs being measured (Bolt & Lall, 2003; de la Torre, 2008). Consequently, the three-dimensional model was chosen to more effectively align with the theoretical foundation of the test and to maximize its practical advantages (Reise & Yu, 1990).

In determining the parameter ranges for the model, relevant literature was carefully reviewed (Baker, 2001; Feinberg & Rubright, 2016; Reise & Yu, 1990). Accordingly, ability parameters were generated from a normal distribution with a mean of 0 and a standard deviation of 1 (see Table 1), which is commonly applied in IRT models.

Table 1. Simulation Design

|  |  | Levels | # Levels |
|---|---|---|---|
| Parameters | Theta ($\theta$) | $\theta \sim N(0, 1)$ |  |
|  | a1 | $U \sim (0.44, 0.75)$ |  |
|  | a2 | $U \sim (0.58, 0.98)$ |  |
|  | a3 | $U \sim (0.75, 1.33)$ |  |
|  | b1 | $U \sim (0.67, 1.34)$ |  |
|  | b2 | b1 - $U \sim (0.67, 1.34)$ |  |
|  | b3 | b2 - $U \sim (0.67, 1.34)$ |  |
| Conditions | Test Length | 20, 40 | 2 |
|  | Interdimensional Correlations | 0.3, 0.7 | 2 |
| # Conditions |  |  | 2x2=4 |
| Replication |  |  | 100 |
| # Datasets |  |  | 4x100=400 |

Note. Theta = ability; a1, a2, a3 = item discrimination parameters; b1, b2, b3 = item difficulty parameters; N = normal distribution; U=uniform distribution

While the item discrimination (a) and difficulty (b) parameters are theoretically defined in the range of (-∞, +∞) in IRT models, practical applications often constrain parameter a between -2.80 and +2.80 and parameter b between -3 and +3. In this study, a and b parameters were determined from a uniform distribution, as shown in Table 1. For items with low discrimination, the parameters were selected from $U \sim (0.44, 0.75)$; for items with moderate discrimination, from $U \sim (0.58, 0.98)$; and for items with high discrimination, from (0,75, 1,33). The item difficulty parameter "b1" was selected from the U ~ (0.67, 1.34) distribution, while the subsequent b parameters were determined by decreasing each of them from the U ~ (0.67, 1.34) distribution starting from b1, respectively. This method is based on the properties of GRM and references in the existing literature (Bulut & Sunbul, 2017; Jiang et al., 2016).

In MC studies, it is suggested that more accurate item parameter estimates can be achieved in tests comprising 20 or more items and with sample sizes of 500 or more (De Ayala, 2009). Based on this premise, the number of items in this study was set at 20 (short) and 40 (long), while the sample size was fixed at 2000 participants. Simulation studies reveal that correlations between dimensions typically range from 0 to 0.9 (Bolt & Lall, 2003; de la Torre, 2009; de la Torre & Hong, 2010; de la Torre & Patz, 2005; Guo & Choi, 2023; Jiang et al., 2016; Kuo & Sheng, 2015; Kuo & Sheng, 2016; Yao & Boughton, 2007). In this study, the correlations between dimensions were set at 0.3 and 0.7 to reflect moderate and relatively strong relationships, respectively. A primary rationale for limiting the correlation values to 0.3 and 0.7 is that these values sufficiently represent low and high correlation scenarios, as supported by previous studies (Bolt & Lall, 2003; de la Torre, 2008; Jiang et al., 2016). Additionally, these values provide sufficient variability for the study's objectives while avoiding unnecessary complexity that could hinder the interpretability of the simulation results. The number of replications was set at 100, resulting in the generation of 100 different datasets for each condition. Therefore, 2 (test lengths) x 2 (interdimensional Correlation) x 100 (replication) = 400 datasets were generated.

*Data Generation*

The *'mvtnorm'* (Genz & Bretz, 2009), *'mirt'* (Chalmers, 2012), and *'rlist'* (Ren, 2016) packages were utilized during the data generation and storage process. The data generation steps and following estimations in this section were carried out considering the condition of 20 items and 0.3 interdimensional correlation. The *'install.packages'* function was employed to download and install the required packages in the R environment, after which the packages were loaded into the environment using the *'library'* function.

```
# Install and load necessary packages
install.packages ("mvtnorm")
library (mvtnorm)
install.packages ("mirt")
library (mirt)
install.packages ("rlist")
library (rlist)
```

After the packages were loaded, the *'set.seed'* function was used to specify the 'seed' for the random number generator. The advantage of this is to ensure the reproducibility of the findings. By using the same *'seed'*, the same parameters or data sets can be obtained each time. After determining the seed of the study (1234), empty lists were created to save the item parameters to be generated with 100 replications. For example, *a1* parameters to be produced with *'a1=list()'* will be saved in this list variable. Similarly, *'a'* and *'b'* lists have been created in which *a1, a2, a3* and *b1, b2, b3* parameters to be produced for 3 dimensions will be kept together. *'Theta'* represents the list variable where the generated ability scores will be stored, while *'datasets'* is the list variable designated for storing the generated datasets. In the next stage, a *'for'* loop was created to generate 100 datasets for each condition. Since the item discrimination and difficulty parameters were generated using a uniform distribution, *'runif'* function was employed. The *a1* parameters for the first 7 items were loaded into Dimension 1 with random values ranging between 0.44 and 0.75, and *'rep(0,13)'* ensures that the value of the next 13 items is 0 and not loaded into Dimension 1. It should be noted that in the generation of *a2* parameters, the values of the first 7 and last 6 items are 0 and are not loaded into Dimension 2. Similarly, in the generation of a3 parameters, the value of the first 14 items is 0 and does not load on Dimension 3. After the item discriminations for the three dimensions were turned into matrices with *'as.matrix'*, the parameters were combined into a single matrix using the *'cbind'* function. In 20x3 matrix, the columns show the dimensions respectively.

```r
set.seed(1234)
# Initialize lists to store simulation results
a1=list()
a2=list()
a3=list()
a=list()
b1=list()
b2=list()
b3=list()
b=list()
Theta=list()
datasets=list()

# Start simulation loop
for (i in 1:100) {

# Generate discrimination parameters for 20 items
a1[[i]]<-as.matrix(c(runif(7,0.44, 0.75), rep(0,13)))
a2[[i]]<-as.matrix(c(rep(0,7),runif(7,0.58, 0.98), rep(0,6)))
a3[[i]]<-as.matrix(c(rep(0,14),runif(6,.75, 1.33)))

# Combine discrimination parameters into a single matrix
a[[i]]<-cbind(a1[[i]],a2[[i]],a3[[i]])

# Generate difficulty parameters for 20 items
b1[[i]] <- runif(20,0.67, 1.34)
```

```
b2[[i]] <- b1[[i]]-runif(20,0.67, 1.34)
b3[[i]] <- b2[[i]]-runif(20,0.67, 1.34)

# Combine and sort difficulty parameters
b[[i]] <- as.matrix(cbind(b1[[i]], b2[[i]], b3[[i]]))
b[[i]] <- t(apply(b[[i]], 1, sort, decreasing=TRUE))

# Generate ability parameters for 2000 sample size
Theta[[i]] <- rmvnorm(2000, mean=c(0,0,0), sigma = matrix(c(1,
                   .3,.3,.3, 1,.3,.3,.3, 1),3,3))
# Simulate item response data
datasets[[i]]<-  simdata(a[[i]],  b[[i]],  2000,  Theta=Theta[[i]],
itemtype = 'graded')
}

# Save the list of datasets
list.save(datasets, "D://datasets.rds")
```

In the generation of the three thresholds for each item, the *b1* parameter was assigned random values ranging from 0.67 to 1.34. The *b2* parameter, representing another threshold, was generated by subtracting random values within the same range from the *b1* parameter. Similarly, the *b3* parameter was produced by subtracting random values from the *b2* parameter. Subsequently, the generated thresholds were combined into a single matrix, and the thresholds were sorted in descending order using them '*apply*' and '*sort*' functions. To ensure sorting from largest to smallest, the argument '*decreasing=TRUE*' was specified. Ability parameters were generated with '*rmvnorm*' for samples of 2000 individuals. '*mean=c(0, 0, 0)*' indicates that the mean vector for the three dimensions is 0. The covariance matrix diagonals with '*sigma*' were set as 1 and the other elements were set as 0.3, which is the interdimensional correlation value. In this way, ability parameters for 3 dimensions were generated. For the generation of item response data for individuals, '*simdata*' was used.

*'itemtype= graded'* specifies that the item responses to be generated will follow the 'Graded Response Model'. After generating the datasets with 4 categories (0, 1, 2, 3), the *'list.save'* function can be used to save them. At this stage, after specifying the target directory for saving, the list variable *(datasets)* can be saved with the *'.rds'* extension. The saved *'.rds'* file contains 100 different datasets.

*Parameter Estimation*

Before conducting parameter estimations on the datasets, a multidimensional item factor model was defined using the *'mirt.model'* function. For datasets containing 20 items, the first 7 items were loaded onto the first factor (F1), the next 7 items onto the second factor (F2), and the final 6 items onto the third factor (F3). For datasets containing 40 items, the first 13 items were assigned to F1, the next 13 to F2, and the final 14 to F3. At this stage, it is assumed that the items are evenly distributed across the three factors.

```
# Define the model for 20 items
model <- mirt.model('F1=1-7
                    F2=8-14
                    F3=15-20')

# Fit the models
mirt_model <- sapply(1:100, function(i) {
  mirt(data = datasets[[i]], model, SE = TRUE, verbose = FALSE)
}, simplify = FALSE)

# Extract item parameters for each model
parameters <- sapply(1:100, function(i) {
  as.data.frame(coef(mirt_model[[i]], simplify = TRUE)$items)
}, simplify = FALSE)
```

The *'mirt'* function was utilized to fit the datasets to the defined multidimensional model. The model argument within the function specifies the factor structure defined in the previous step. The option *'SE=TRUE'* was used to obtain standard errors for the parameter estimates, while *'verbose=FALSE'* suppressed detailed information about the model-fitting process. If detailed information is required, this setting can be changed to *TRUE*. The use of *'simplify=FALSE'* ensures that the output of the sapply() function is returned as a list. The duration of the model estimation process varies depending on the number of conditions, the dimensionality of the model, and the computational power of the machine used. Ultimately, the model was fitted to all 100 datasets.

The *'coef'* function was used for the estimation of item parameters, with the *'$items'* argument indicating a focus on item-related parameters. The extracted parameters were converted into a data frame for easier handling using the *'as.data.frame'* function. In this way, the estimated parameters for the three-dimensional MIRT model across 100 datasets were stored in the *'parameters'* variable.

*Bias and RMSE Calculations*

The calculation of bias and RMSE values for the estimated parameters was based on the following formulas:

$$bias = \frac{\sum_{i=1}^{K}(\hat{X}_i - X_i)}{K} \quad (1)$$

$$RMSE = \sqrt{\frac{\sum_{i=1}^{K}(\hat{X}_i - X_i)^2}{K}} \quad (2)$$

In both equations, K is the number of items, $\hat{X}_i$ is the parameter estimation for item i (i=1, 2, ..., K) and $X_i$ is the actual parameter estimation for item i. When calculating bias and RMSE values, the names of the item parameters (a1, a2, a3, b1, b2, b3) and the names of the estimated item parameters (a1, a2, a3, d1, d2, d3) should be considered. Here, 'd' refers to the item difficulty parameter. The bias for a1 parameters is calculated with *'sum(parameters[[i]]$a1 – a1[[i]]) / 7'* shown below. *'parameters[[i]]$a1'* indicates the estimated a1 parameters and *'a1[[i]]'* indicates the actual a1 parameters. The difference of the values related to the parameters are summed and divided by the total number of a1 parameters (7). When calculating bias and RMSE for the a2 and a3 parameters, the denominators should be set to 7 and 6, respectively, and the parameter names in the code must be changed to a2 and a3.

```
# Calculate bias for a1 parameter
bias_a <- sapply(1:100, function(i) {
  sum(parameters[[i]]$a1 - a1[[i]]) / 7
})

# Calculate the mean bias and round it to 3 decimal places
bias_a <- round(mean(bias_a), 3)

# Display the result
bias_a

# Calculate RMSE for for a1 parameter
rmse_a <- sapply(1:100, function(i) {
  sqrt(sum((parameters[[i]]$a1 - a1[[i]])^2) / 7)
})

# Calculate the mean RMSE and round it to 3 decimal places
rmse_a <- round(mean(rmse_a), 3)

# Display the result
rmse_a
```

For the item difficulty parameter b1, while calculating bias, the estimated parameters are represented as parameters[[i]]$d1, and the actual parameters are represented as b1[[i]]. The differences between the values of these parameters are summed and divided by the total number of b1 parameters (20). For the b2 and b3 parameters, it is sufficient to change only the parameter names in the code. *'round(mean(bias), 3)'* calculates the average bias value across the 100 datasets and rounds the result to three decimal places using the *'round'* function. Similarly, RMSE values are calculated and averaged below.

```
# Calculate bias for b1 parameter
bias_b <- sapply(1:100, function(i) {
  sum(parameters[[i]]$d1 - b1[[i]]) / 20
})

# Calculate the mean bias and round it to 3 decimal places
bias_b <- round(mean(bias_b), 3)

# Display the result
bias_b

# Calculate RMSE for for b1 parameter
rmse_b <- sapply(1:100, function(i) {
  sqrt(sum((parameters[[i]]$d1 - b1[[i]])^2) / 20)
})

# Calculate the mean RMSE and round it to 3 decimal places
rmse_b <- round(mean(rmse_b), 3)

# Display the result
rmse_b
```

The bias averages calculated for the item parameters considering the simulation conditions are shown in Table 2. At this point, in the case where the interdimensional correlation is 0.3, it is seen that the bias averages decreased for a2, a3, b1 and b3 parameters with an increase in test length, while the bias for a1 and b2 parameters remained unchanged. When the interdimensional correlation was 0.7, the bias averages increased for a1, decreased for a2 and b1, and remained unchanged for a3, b2 and b3 with an increase in test length. Moreover, it was observed that an increase in the interdimensional correlation resulted in a decrease in the bias average for the a1 parameter when the test length was 20, while no change was observed for the other parameters. When the test length was 40, the increase in interdimensional correlation led to an increase in the bias averages for the a3 and b3 parameters, whereas no change was observed for the other parameters.

Table 2. Average bias for item parameters

| Test Length | Interdimensional Correlation | a1 | a2 | a3 | b1 | b2 | b3 |
|---|---|---|---|---|---|---|---|
| 20 | 0.3 | .002 | -.002 | .003 | -.001 | .000 | -.001 |
| 20 | 0.7 | .001 | -.002 | .003 | -.001 | .000 | -.001 |
| 40 | 0.3 | .002 | -.001 | -.002 | .000 | .000 | .000 |
| 40 | 0.7 | .002 | -.001 | -.003 | .000 | .000 | -.001 |

Note. a1, a2, a3 = item discrimination parameters; b1, b2, b3 = item difficulty parameters

When the RMSE averages for item parameters are analyzed, it is seen that lower RMSE averages are obtained with the increase in test length from 20 to 40 for a parameters (see Table 3). However, for the b parameters, the RMSE averages did not show a noticeable change with the increase in test length. In addition, it was determined that the increase in the interdimensional correlation did not cause a considerable change in the RMSE averages for the parameters.

Table 3. Average RMSE for item parameters

| Test Length | Interdimensional Correlation | a1 | a2 | a3 | b1 | b2 | b3 |
|---|---|---|---|---|---|---|---|
| 20 | 0.3 | .069 | .069 | .074 | .057 | .050 | .057 |
| 20 | 0.7 | .069 | .068 | .073 | .057 | .051 | .057 |
| 40 | 0.3 | .062 | .059 | .061 | .056 | .050 | .056 |
| 40 | 0.7 | .061 | .059 | .061 | .056 | .050 | .056 |

Note. a1, a2, a3 = item discrimination parameters; b1, b2, b3 = item difficulty parameters

*Organizing and Visualizing Findings*

If the findings are to be visualized, the process can begin by preparing an Excel file to store the results. As shown in Table 4, the first two columns represent the simulation conditions, specifically *'Test_Length'* and *'Dimension'*. The remaining columns include the estimated item parameters *'Parameters'* and the calculated average *'Bias'* and *'RMSE'* values. Once the findings are organized into a table, they can be saved with the *'.xlsx'* extension. In this example, the results were saved to the *'D'* drive under the name *'results.xlsx'*.

Table 4. Excel table containing the findings

| Test_Length | Dimension | Parameters | Bias | RMSE |
|---|---|---|---|---|
| 20 | 0.3 | a1 | 0.002 | 0.069 |
| 20 | 0.3 | a2 | -0.002 | 0.069 |
| 20 | 0.3 | a3 | 0.003 | 0.074 |
| 20 | 0.3 | b1 | -0.001 | 0.057 |
| 20 | 0.3 | b2 | 0.000 | 0.050 |
| 20 | 0.3 | b3 | -0.001 | 0.057 |
| 20 | 0.7 | a1 | 0.001 | 0.069 |
| 20 | 0.7 | a2 | -0.002 | 0.068 |
| 20 | 0.7 | a3 | 0.003 | 0.073 |
| 20 | 0.7 | b1 | 0.000 | 0.057 |
| 20 | 0.7 | b2 | 0.000 | 0.051 |
| 20 | 0.7 | b3 | 0.000 | 0.057 |
| 40 | 0.3 | a1 | 0.002 | 0.062 |
| 40 | 0.3 | a2 | -0.001 | 0.059 |
| 40 | 0.3 | a3 | -0.002 | 0.061 |
| 40 | 0.3 | b1 | 0.000 | 0.056 |
| 40 | 0.3 | b2 | 0.000 | 0.050 |
| 40 | 0.3 | b3 | 0.000 | 0.056 |
| 40 | 0.7 | a1 | 0.002 | 0.061 |
| 40 | 0.7 | a2 | -0.001 | 0.059 |
| 40 | 0.7 | a3 | -0.003 | 0.061 |
| 40 | 0.7 | b1 | 0.000 | 0.056 |
| 40 | 0.7 | b2 | 0.000 | 0.050 |
| 40 | 0.7 | b3 | -0.001 | 0.056 |

Note. a1, a2, a3 = item discrimination parameters; b1, b2, b3 = item difficulty parameters

To import the saved Excel document into the R environment and to create graphs, *'readxl'* and *'ggplot2'* packages must be installed and then loaded. At this point, the *'results.xlsx'* file saved with the *'read_excel'* function was extracted into the R environment and saved in the *'table'* variable. Although the variables in the first three columns of the Excel file contain numerical values, these variables are factor variables. Therefore, the names of these variables were first saved in the variables *'names1'*, *'names2'* and *'names3'*, and subsequently converted into factors using the *'lapply'* function. With *'levels'*, the levels of each factor and their order were determined. *'Bias'* and *'RMSE'* variables are numeric variables. Therefore, with *'as.numeric'* it is indicated that these variables are numeric variables.

```
# Install and load the necessary packages
install.packages("readxl")
library(readxl)
install.packages("ggplot2")
library(ggplot2)

# Load the Excel file into a data frame
table<-read_excel("D://results.xlsx")

# Define column names
names1<-c('Test_Length')
names2<-c('Dimension')
names3<-c('Parameters')

# Convert 'Test_Length', 'Dimension', and 'Parameters' columns to
factor variables with specified levels
table[,names1]<-lapply(table[,names1], factor, levels=c("20", "40"))
table[,names2]<-lapply(table[,names2], factor, levels=c("0.3",
"0.7"))
table[,names3]<-lapply(table[,names3], factor, levels=c("a1", "a2",
"a3", "b1", "b2", "b3"))
```

```
# Convert 'Bias' and 'RMSE' columns to numeric variables
table$Bias<-as.numeric(table$Bias)
table$RMSE<-as.numeric(table$RMSE)
```

*'ggplot'* is a versatile function that allows the creation of a wide variety of high-quality graphs. In this case, the *'data'* argument is used to link the plot to the table containing the results. Within the plot aesthetics mapping, *'x = factor(Dimension)'* indicates that the *'Dimension'* variable will be displayed on the x-axis, while *'y = Bias'* specifies that the Bias variable will be shown on the y-axis. The *'Parameters'* variable is included within shape, color, and group, which determines that the shapes, colors, and grouping within the plot will be based on this variable.

The size of the plot shapes can be adjusted with *'geom_point(size = 5)'*, while the line thickness can be controlled using *'geom_line(linewidth = 1)'*. Using *'facet_grid'*, separate plots (facets) are created for the *'test length'* variable, and custom labels for each facet can be added with *'labeller'*. The *'scale_shape_manual'* and *'scale_color_manual'* functions allow for customizing the shapes and colors of the plot elements. To label the x and y axes, the *'labs'* function is used. The y-axis limits can be set with *'coord_cartesian(ylim = c(-0.01, 0.01))'*, and it is important to determine these limits based on the value ranges of the obtained results.

```
# Create the ggplot object for the plot
plot_bias <- ggplot(data = table, aes(x = factor(Dimension), y =
Bias, shape = factor(Parameters), color = factor(Parameters), group
= factor(Parameters))) +
    geom_point(size = 5) +
    geom_line(linewidth = 1) +
      facet_grid(. ~ Test_Length, labeller = labeller(Test_Length =
        c("20" = "Test Length = 20", "40" = "Test Length = 40"))) +
```

```
    scale_shape_manual(values = c(15, 16, 17, 18, 22, 23),
                       name = "Parameters") +
    scale_color_manual(values = c("black", "red", "green", "blue",
          "purple","cyan"), name = "Parameters") +
    labs(x = "Interdimensional Correlation", y = "Bias") +
    coord_cartesian(ylim = c(-0.01, 0.01)) +
theme(
    axis.title = element_text(size = 18),
    axis.text = element_text(size = 16, color = "black"),
    strip.text = element_text(size = 16, face = "bold"),
    legend.text = element_text(size = 16),
    legend.title = element_text(size = 18)
  )

# Display the plot
plot_bias
```

In the *'theme'* section, the visual properties of the plot, such as text size and color, can be customized. Specifically, *'axis.title'* is used to modify the labels for the x and y axes, *'axis.text'* adjusts the numerical values on the axes, *'strip.text'* customizes the labels for the facets created by the test length variable, and *'legend.text'* and *'legend.title'* are used to modify the text size, color, and other properties of the legend. After executing the relevant code for bias and RMSE, the resulting plots are presented in Figure 1 and Figure 2.

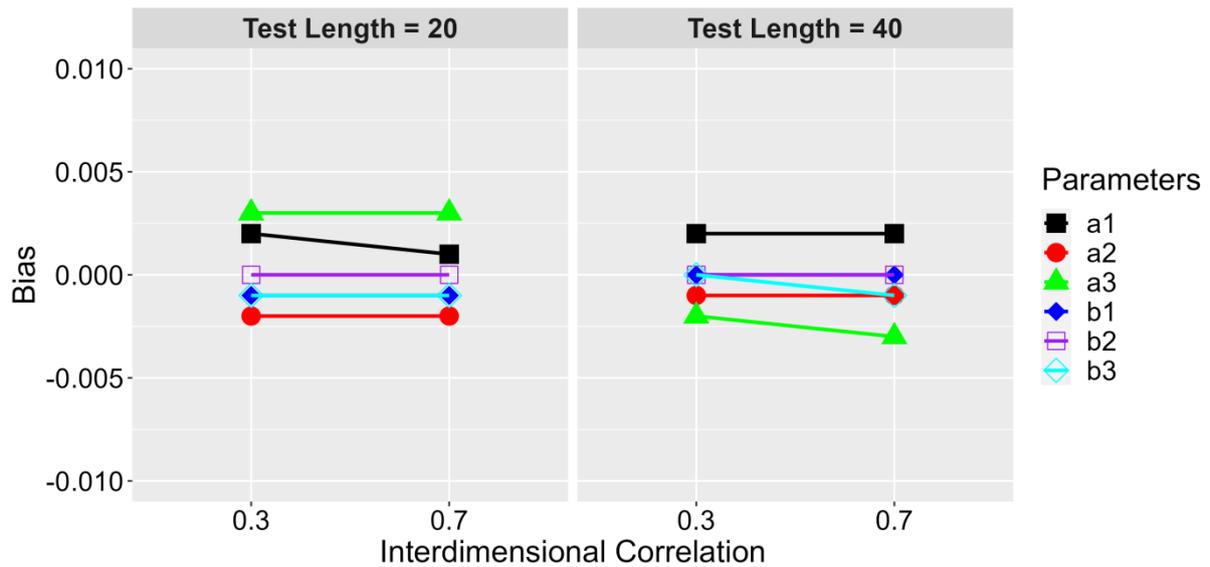

**Figure 1.** Average bias for item parameters
Note. a1, a2, a3 = item discrimination parameters; b1, b2, b3 = item difficulty parameters

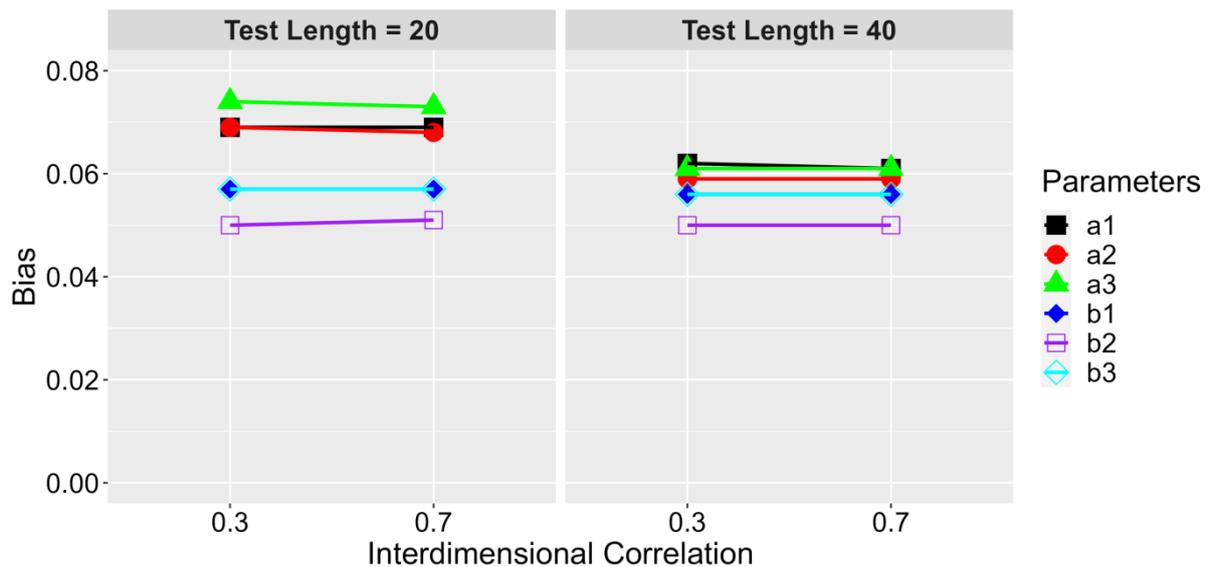

**Figure 2.** Average RMSE for item parameters
Note. a1, a2, a3 = item discrimination parameters; b1, b2, b3 = item difficulty parameters

For the plots to be saved to the computer drive with a pixel resolution of 300 dpi, *'ggsave'* was used. *'bias.png'* indicates the name of the plot to be saved, while *'plot = plot_bias'* refers to the *'plot_bias'* variable created with ggplot. *'width'* and *'height'* indicate the width and height of the plot, and *'units'* indicate that this numerical information will be considered in cm. The *"path"* specifies the directory where the plot will be saved, while *"dpi"* determines the resolution of the generated image.

```
# Save the plot as a PNG image with specified dimensions and 
resolution
ggsave("bias.png", plot = plot_bias, width = 25, height = 12, path = 
"D://", units = "cm", dpi = 300)
```

**Discussion**

Data generation and item recovery processes are becoming increasingly important in educational sciences, particularly in areas such as test development, measurement and evaluation, and psychometric analyses. In this context, simulation-based data generation has become an important tool for overcoming the challenges of empirical data collection and for examining model performance under varying conditions in educational sciences. Furthermore, simulation studies not only overcome the difficulties of empirical data collection but are also preferred for creating controlled and experimental conditions for method comparisons and similar analyses. In this regard, simulations allow for systematic testing of different conditions, enabling a more consistent examination of parameter effects.

Within this framework, item recovery studies offer significant advantages for both theoretical and applied research in the field of educational sciences. Theoretically, they provide an opportunity to assess the validity and accuracy of models designed for complex test structures. Practically, they establish a foundation for examining the effects of varying conditions on parameter estimation and for developing methods to enhance estimation accuracy. Additionally, they enable researchers to anticipate potential issues and optimize model development processes prior to transitioning to the real data collection phase.

Building on these foundations, this study aims to provide a detailed presentation of the item recovery process in MGRM using R. The R codes, simulation design, and parameter descriptions presented in this work serve as a comprehensive guide for researchers on data generation, parameter estimation, bias and RMSE calculations, and result visualization within the scope of item recovery in educational sciences. Furthermore, the methods outlined here are expected to benefit researchers not directly engaged in item recovery, particularly for tasks such as data generation, parameter estimation, and visualization.

This study underscores the practicality of using R for item recovery in the context of MGRM, providing researchers with a comprehensive and robust toolkit for conducting similar analyses. Future research can build on the approaches presented here, contributing to advancements in both psychometric research and applied studies. In this regard, the robustness of item recovery could be further evaluated under broader scenarios, such as incorporating different parameter ranges and distributions, varying test lengths, or exploring alternative multidimensional structures. More extreme conditions, including the presence of missing data or non-normally distributed latent traits, could also be examined in future studies. While R is highly user-friendly and accessible, computational efficiency becomes a significant consideration when handling larger datasets or higher-dimensional models. This aspect should be carefully addressed when defining simulation conditions. Furthermore, the integration of MGRM into adaptive testing frameworks warrants exploration, with a focus on its scalability and practical benefits in operational settings.

**Declaration of Conflicting Interests**

The authors declared no potential conflicts of interest with respect to the research, authorship, and/or publication of this article.

**Funding**

The authors received no financial support for the research, authorship, and/or publication of this article.

**ORCID**

Yeşim Beril Soğuksu 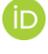 https://orcid.org/0009-0004-0870-4974

Ayşe Bilicioğlu Güneş 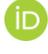 https:// orcid.org/0000-0002-1603-8631

Hatice Gürdil 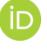 https:// orcid.org/0000-0002-0079-3202

**References**

Ackerman, T. (1996). Graphical representation of multidimensional item response theory analyses. *Applied Psychological Measurement*, *20*(4), 311–329.

Ackerman, T.A., Gierl, M.J., & Walker, C.M. (2003). Using multidimensional item response theory to evaluate educational and psychological tests. *Educational Measurement: Issues and Practice, 22*(3), 4-54.

Adams, R. J., Wilson, M. R., & Wang, W.-C. (1997). The multidimensional random coefficients multinomial logit model. *Applied Psychological Measurement, 21*, 1-23.

Alarcon, G. M., Lee, M. A., & Johnson, D. (2023). A Monte Carlo study of IRTree models' ability to recover item parameters. *Frontiers in Psychology, 14,* 1003756.

Andrich, D. (1978). Application of a psychometric rating model to ordered categories which are scored with successive integers. *Applied Psychological Measurement, 2*(4), 581–594.

Baker, F. B. (2001). *The basics of item response theory* (2nd ed.). College Park, (MD): ERIC Clearinghouse on Assessment and Evaluation.

Bilicioglu Gunes, A. (2024). *Madde parametre kaymasinin bulundugu durumlarda test esitlemede farkli kalibrasyon yontemleriyle elde edilen kestirimlerin karsilastirilmesi (A comparison of estimates obtained with different calibration methods in test equating in the presence of item parameter drift).* [Doctoral dissertation, Ankara University]. YOK Tez.

Birnbaum, A. (1968). Some latent trait models and their use in inferring an examinee's ability. In F.M. Lord & M.R. Novick, *Statistical theories of mental test scores* (pp. 392-479). MA: Addison-Wesley.

Bock, R. D. (1972). Estimating item parameters and latent ability when responses are scored in two or more nominal categories. *Psychometrika, 37,* 29-51.

Bolt, D. M. ve Lall, V. F. (2003). Estimation of compensatory and noncompensatory multidimensional item response models using Markov chain Monte Carlo. *Applied Psychological Measurement*, *27*(6), 395-414.

Bulut, O., & Sünbül, Ö. (2017). R programlama dili ile madde tepki kuramında monte carlo simülasyon çalışmaları. *Journal of Measurement and Evaluation in Education and Psychology, 8*(3), 266-287.

Carle, A., Jaffee, D., Vaughan, N., & Eder, D. (2009). Psychometric properties of three new National Survey of Student Engagement based engagement scales: An item response theory analysis. *Research in Higher Education, 50*(8), 775–794.

Chalmers, R. P. (2012). mirt: A multidimensional item response theory package for the R environment. *Journal of Statistical Software, 48*(6), 1–29. https://doi.org/10.18637/jss.v048.i06

De Ayala, R. J. (2009*). The theory and practice of item response theory.* New York: The Guilford Press.

de la Torre, J. (2008). Multidimensional scoring of abilities: The ordered polytomous response case. *Applied Psychological Measurement*, *32*(5), 355-370.

de la Torre, J. (2009). DINA model and parameter estimation: A didactic. *Journal of educational and behavioral statistics*, *34*(1), 115-130.

de la Torre, J., Hong, Y., & Deng, W. (2010). Factors affecting the item parameter estimation and classification accuracy of the DINA model. *Journal of Educational Measurement*, *47*(2), 227-249.

de la Torre, J. ve Patz, R.J. (2005). Making the most of what we have: a practical application of multidimensional item response theory ın test scoring. *Journal of Educational and Behavioral Statistics, 30*(3), 295-311.

Embretson, S. E., & Reise, S. P. (2000). *Item response theory for psychologists.* Lawrence Erlbaum Associates Publishers.

Feinberg, R. A., & Rubright, J. D. (2016). Conducting simulation studies in psychometrics. *Educational Measurement: Issues and Practice, 35* (2), 36-49.

French, B. F., & Vo, T. T. (2020). Differential item functioning of a truancy assessment. *Journal of Psychoeducational Assessment*, *38*(5), 642-648.

Furlow C. F., Ross T. R., Gagné P. (2009). The impact of multidimensionality on the detection of differential bundle functioning using the Simultaneous Item Bias Test. *Applied Psychological Measurement*, 33, 441-464.

Genz, A., & Bretz, F. (2009). *Computation of multivariate normal and t probabilities* (Vol. 195). Springer Science & Business Media.

Guo, W., & Choi, Y. J. (2023). Assessing dimensionality of IRT models using traditional and revised parallel analyses. *Educational and Psychological Measurement*, *83*(3), 609-629.

Gurdil, H., & Demir, E. (2024). The use of multidimensional item response theory estimations in controlling differential item functioning. *Measurement: Interdisciplinary Research and Perspectives,* 1-14.

Gurdil, H., Soguksu, Y. B., Salihoglu, S., & Coskun, F. (2024). Integrating AI in educational measurement: ChatGPT's efficacy in item response theory data generation. *arXiv preprint arXiv:2402.01731*.

Hallgren, K. A. (2013). Conducting simulation studies in the R programming environment. *Tutorials in Quantitative Methods for Psychology, 9* (2), 43-60.

Hambleton, R. K., Swaminathan, H., & Rogers, H. (1991). *Fundamentals of item response theory.* Newbury Park CA: Sage.

Han, K. C. T., & Paek, I. (2014). A review of commercial software packages for multidimensional IRT modeling. *Applied Psychological Measurement*, *38*(6), 486-498.

Hartig, J., & Höhler, J. (2009). Multidimensional IRT models for the assessment of competencies. *Studies in Educational Evaluation, 35,* 57- 63.

Harwell, M., Stone, C. A., Hsu, T., & Kirisci, L. (1996). Monte Carlo studies in item response theory. *Applied Psychological Measurement, 20* (2), 101-125.

Hattie, J. (1985). Methodology review: Assessing unidimensionality of tests and items. *Applied Psychological Measurement, 9*, 139–164.

Jiang, C., Li, D., Wang, D., & Zhang, L. (2016). Quantification and assessment of changes in ecosystem service in the Three-River Headwaters Region, China as a result of climate variability and land cover change. *Ecological Indicators*, *66*, 199-211.

Jiang, S., Wang, C., & Weiss, D. J. (2016). Sample size requirements for estimation of item parameters in the multidimensional graded response model. *Frontiers in Psychology, 7*, 109. https://doi.org/10.3389/fpsyg.2016.00109

Kehinde, O. J., Dai, S., & French, B. (2022). Item parameter estimations for multidimensional graded response model under complex structures. *Frontiers in Education, 7*. https://doi.org/10.3389/feduc.2022.947581

Kroese, D. P., Brereton, T., & Taimre, T. (2014). Why the Monte Carlo method is so important today. *WIREs Comput Stat, 6,* 386–392.

Kuo, T. C., & Sheng, Y. (2015). Bayesian estimation of a multi-unidimensional graded response IRT model. *Behaviormetrika*, *42*(2), 79-94.

Kuo, T. C., & Sheng, Y. (2016). A comparison of estimation methods for a multi-unidimensional graded response IRT model. *Frontiers in Psychology*, *7*, 880.

Lord, F. M. (1952). *A theory of test scores* (Psychometric Monograph No. 7). Iowa City, IA: Psychometric Society, 35.

Lord, F.M. (2012). *Applications of item response theory to practical testing problems*. Routledge. https://doi.org/10.4324/9780203056615

Meijer, R. R., & Tendeiro, J. N. (2018). Unidimensional item response theory. In P. Irwing, T. Booth, & D. J. Hughes (Eds.), *The Wiley handbook of psychometric testing: A* *multidisciplinary reference on survey, scale and test development* (pp. 413–443). Wiley Blackwell. https://doi.org/10.1002/9781118489772.ch15

Muraki, E., & Carlson, J. E. (1995). Full-information factor analysis for polytomous item responses. *Applied Psychological Measurement, 19*(1), 73-90. https://doi.org/10.1177/014662169501900109

Reckase, M. D. (2009). *Multidimensional item response theory (Statistics for social and behavioral sciences)*. New York: Springer.

Reise, S. P., & Yu, J. (1990). Parameter recovery in the graded response model using MULTILOG. *Journal of Educational Measurement*, *27*(2), 133-144.

Ren, K. (2016). Rlist: A toolbox for non-tabular data manipulation. *R package version 0.4*, *6*.

Samejima, F. (1969). Estimation of latent ability using a response pattern of graded scores. (Psychometric Monograph No. 17). Richmond, VA: Psychometric Society. Retrieved from http://www.psychometrika.org/journal/online/MN17.pdf

Sharkness, J., & DeAngelo, L. (2011). Measuring student involvement: A comparison of classical test theory and item response theory in the construction of scales from student surveys. *Research in Higher Education*, *52*, 480-507.

Sijtsma, K., & Junker, B.W. (2006). Item response theory: Past performance, present developments and future expectations. *Behaviormetrika, 1*, 75-102.

Spearman, C. (1904). The proof and measurement of association between two things. *American Journal of Psychology, 15*, 72-101.

Strachan, T., Ip, E., Fu, Y., Ackerman, T., Chen, S. H., & Willse, J. (2020). Robustness of projective IRT to misspecification of the underlying multidimensional model. *Applied Psychological Measurement*, *44*(5), 362-375.

Thissen, D.M. (1976). Information in wrong responses to the raven progressive matrices. *Journal of Educational Measurement, 13*, 201-214.

Uttaro, T., & Lehman, A. (1999). Graded response modeling of the Quality of Life Interview. *Evaluation and Program Planning*, *22*(1), 41-52.

Walker, C. M., & Gocer Sahin, S. (2016). Using a multidimensional IRT framework to better understand differential item functioning (DIF): A tale of three DIF detection procedures. *Educational and Psychological Measurement, 77*(6), 945-970.

Xu, J., Paek, I., & Xia, Y. (2017). Investigating the behaviors of M 2 and RMSEA2 in fitting a unidimensional model to multidimensional data. *Applied Psychological Measurement*, *41*(8), 632-644.

Yao, L., & Schwarz, R. D. (2006). A multidimensional partial credit model with associated item and test statistics: An application to mixed-format tests. *Applied Psychological Measurement, 30*(6), 469–492.

Zhang, J. (2012). Calibration of response data using MIRT models with simple and mixed structures. *Applied Psychological Measurement*, *36*(5), 375-398.